\documentclass[allclo]{FBSart}
\usepackage{amsfonts}
\usepackage{amssymb}
\usepackage{epsfig}

\title{{\em Ab initio} no-core shell model and microscopic reactions: recent achievements}
\author{S. Quaglioni \thanks{\textit{E-mail address:} 
quaglioni1@llnl.gov}, P. Navr{\'a}til}
\institute{Lawrence Livermore National Laboratory, L-414, P.O. Box 808, Livermore, CA 94551, USA}

\runningauthor{S.\,Quaglioni and P. Navr{\'a}til}
\runningtitle{Ab initio no-core shell model and microscopic reactions: recent achievements}
\begin{document}

\maketitle

\begin{abstract}
We report on recent microscopic calculations of reaction properties based upon the nuclear structure of the {\em ab initio} no-core shell model (NCSM).
\end{abstract}
\section{Introduction}
We have explored two avenues for the {\em ab initio} calculations of nuclear-reaction properties starting from the nuclear structure of the NCSM~\cite{ncsm}: i) the Lorentz integral transform (LIT) approach~\cite{lit}, and ii) the resonating-group method (RGM)~\cite{rgm}. 
In Sect.~\ref{LIT}, we highlight a recent application of the NCSM/LIT approach to the {\em ab initio} calculation~\cite{He4photo} of the $^4$He total photo-absorption cross section. In Sect.~\ref{RGM}, we outline the NCSM/RGM approach and present its application to neutron-$^4$He scattering.
\section{Application of chiral effective field theory two- plus three-nucleon forces to the $^4$He photodisitegration\label{LIT}}
We performed an {\em ab initio} calculation~\cite{He4photo} of the $^4$He total photo absorption cross section in unretarded dipole approximation, using the high quality NN potential at the fourth order (N$^3$LO) in the chiral effective field theory ($\chi$EFT) expansion of Ref.~\cite{entem}, and the NNN interaction at the highest order presently available (N$^2$LO)~\cite{nnn}.  The two low-energy constants that parameterize the short-range NNN interaction were selected as discussed in Ref.~\cite{lecs}. The microscopic treatment of the continuum problem is achieved by means of the LIT method, applied within the NCSM approach. 
\begin{figure}[hbt]
\begin{center}
\epsfig{file=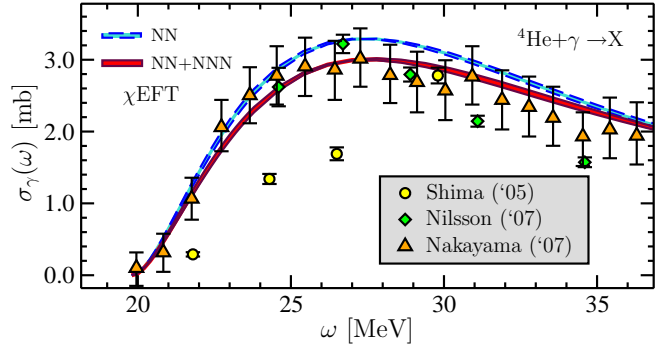,width=250pt}
\caption[]{The $^4$He photo-absorption cross section as a function of the excitation energy $\omega$. Present results with $\chi$EFT interactions compared to the most recent experiments~\cite{expphoto}.\label{He4cs}}
\end{center}
\end{figure}
Accurate convergence of the NCSM expansions is reached using three-body effective interactions for both the NN and NNN terms of the potential. The results, presented in Fig.~\ref{He4cs}, show a peak around the excitation energy of $\omega=27.8$ MeV, with peak-height mildly sensitive to the  NNN force. The experimental situation in the near-threshold region is controversial~\cite{expphoto}:  we find an overall good agreement with the photo-disintegration data from bremsstrahlung photons of Nilsson et al. and the indirect measurements of Nakayama et al.
\section{Augmenting the NCSM by the RGM technique\label{RGM}}
For reactions involving two nuclear fragmets, the RGM many-body wave function is expanded in terms of binary-cluster channel wave functions 
\begin{equation}\label{RGM_wave}
\Psi^{(A)}=\sum_{\nu}\hat{\mathcal A}
\left[\psi_{1\nu}^{(A-a)}\psi_{2\nu}^{(a)}\varphi_{\nu}(\vec{r}_{A-a,a})\right]
=\sum_{\nu}\int d\vec{r}\,\varphi_{\nu}(\vec{r}\,)\,\hat{\mathcal A}\,\Phi_{\nu\vec{r}}^{(A-a,a)}\,,
\end{equation}
with
\begin{equation}\label{RGM_basis}
\Phi_{\nu\vec{r}}^{(A-a,a)}=\psi_{1\nu}^{(A-a)}\psi_{2\nu}^{(a)}\delta(\vec{r}\,-\vec{r}_{A-a,a})\,.
\end{equation}
Here, $\hat{\mathcal A}$ is the antisymmetrizer accounting for the exchanges of nucleons 
between the two clusters (which are already antisymmetric with respect to exchanges
of internal nucleons). 
The relative-motion wave functions $\varphi_{\nu}$ depend on
the relative distance between the center of masses of the two clusters in channel $\nu$. 
They can be determined by solving the many-body Schr\"{o}dinger
equation in the Hilbert space spanned by the basis functions (\ref{RGM_basis}):
\begin{equation}
H\Psi^{(A)}=E\Psi^{(A)}\longrightarrow\, \sum_{\nu}\int d\vec{r}\,
\left[{\mathcal H}^{(A-a,a)}_{\mu\nu}(\vec{r}\,^\prime,\vec{r}\,)
-E{\mathcal N}^{(A-a,a)}_{\mu\nu}(\vec{r}\,^\prime,\vec{r}\,)\right]\varphi_{\nu}(\vec{r}\,)\,,
\end{equation}
where $H$ denotes the full Hamiltonian of the system, and the two integration kernels are defined as
\begin{eqnarray}
{\mathcal H}^{(A-a,a)}_{\mu\nu}(\vec{r}\,^\prime,\vec{r}\,)
&=&\left\langle\Phi_{\mu\vec{r}\,^\prime}^{(A-a,a)}\left|\hat{\mathcal A}\,H\,
\hat{\mathcal A}\right|\Phi_{\nu\vec{r}}^{(A-a,a)}\right\rangle\,,\\
{\mathcal N}^{(A-a,a)}_{\mu\nu}(\vec{r}\,^\prime,\vec{r}\,)&=&
\left\langle\Phi_{\mu\vec{r}\,^\prime}^{(A-a,a)}\left|\hat{\mathcal A}^2\right|
\Phi_{\nu\vec{r}}^{(A-a,a)}\right\rangle\,.
\end{eqnarray}
\begin{figure}[hbt]
\begin{center}
\epsfig{file=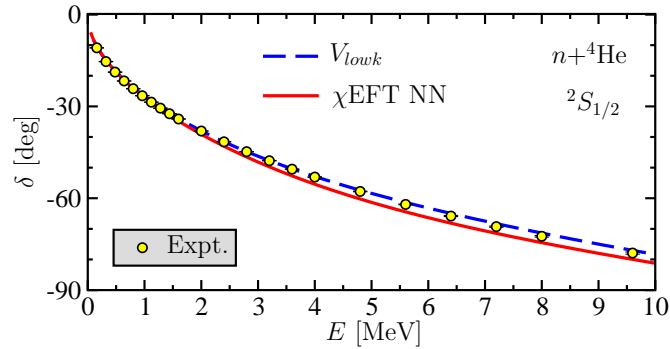,width=250pt}
\caption{The calculated $^2S_{1/2}$ phase shift for the $n+^4$He system as a function of the relative kinetic energy $E$, compared to experiment. The $V_{lowk}$~\cite{vlowk} potential at $\hbar\Omega=18$ MeV in a $16\hbar\Omega$ model space,  and the $\chi$EFT NN interaction~\cite{entem} at  $\hbar\Omega=19$ MeV in a $14\hbar\Omega$ model space were used.\label{figphase}}
\end{center}
\end{figure}
In the NCSM/RGM approach we use {\em ab initio} NCSM wave functions for each of the two clusters, and NCSM effective interactions derived from realistic forces. The most challenging task is to evaluate the Hamiltonian and norm kernels: so far we have developed formalism for cluster states with a single-nucleon projectile and we have applied it to the scattering of low-energy neutrons on $^4$He. In Fig.~\ref{figphase}, we present the converged $^2S_{1/2}$ phase shift for the  $n+^4$He system obtained using the bare $V_{lowk}$~\cite{vlowk} potential, and two-body effective interactions derived from the $\chi$EFT NN potential of Ref.~\cite{entem}. We find an overall good agreement with experiment.
\begin{acknowledge}
This work was performed under the auspices of the U.S. Department of Energy by Lawrence Livermore National Laboratory under Contract DE-AC52-07NA27344. Support from U.S. DOE/SC/NP (Work Proposal Number SCW0498) and the Department of Energy under Grant DE-FC02-07ER41457 is acknowledged. 
\end{acknowledge}
\end{document}